\documentclass[10pt,onecolumn]{article}

\title{Transport Properties of a non-relativistic Delta-Shell Gas with Long 
Scattering Lengths}

\author{SERGEY POSTNIKOV\\
Instituto de Astronom\'{\i}a \\ 
Universidad Nacional Aut\'onoma de M\'exico \\ 
M\'exico D.F. 04510, Mexico\\
and\\
Department of Physics and Astronomy, Ohio University\\
Athens, Ohio 45701, USA\\
sp315503@ohio.edu\\
\\
MADAPPA PRAKASH\\
Department of Physics and Astronomy, Ohio University\\
Athens, Ohio 45701, USA\\
prakash@harsha.phy.ohiou.edu}

\usepackage{mathrsfs}
\usepackage{amsmath}
\usepackage{latexsym}
\usepackage{graphicx}

\begin{document}

\maketitle

\begin{abstract}
The coefficients of diffusion, thermal conductivity, and shear
viscosity are calculated for a system of non-relativistic particles
interacting via a delta-shell potential $V(r)=-v \, \delta(r-R)$ when
the average distance between particles is smaller than $R$.  The roles
of resonances and long scattering lengths including the unitary limit
are examined.  Results for ratios of diffusion to viscosity and
viscosity to entropy density are presented for varying scattering
lengths.
\end{abstract}


\section{Introduction}

Recently, studies of transport properties, particularly viscosities of
interacting particles in a many-body system, have attracted much
attention as observed phenomena in diverse fields of physics (such as
atomic physics and relativistic heavy ions physics) are greatly
influenced by their effects~\cite{ST09}. In atomic physics, the relaxation
of atomic clusters from an ordered state have been
observed~\cite{Kinast05}. In relativistic heavy-ion physics, viscosities
influence the observed elliptic flow versus transverse momentum of
final state hadrons~\cite{HMS09}. In both of these fields, the challenge of
determining the transport properties at varying temperatures and
densities has been vigorously pursued~\cite{ST09}.

In this paper, we wish to study a simple system in which the high
temperature properties of a dilute gas are significantly affected by
the nature of two-body interactions. The two-body interaction chosen
for this study - the delta shell interaction - allows us study the
roles of long scattering lengths, finite-range corrections, and
resonances.  Our study is mainly pedagogical, but several universal
characteristics are revealed. Our results stress the need to perform
analyses beyond what we offer here, particularly with regard to
contributions from higher than two-body interactions.

Gottfried's treatment~\cite{Gottfried66,Gottfried04} of the
quantum mechanics of two particles interacting through a delta-shell
potential $V(r)=-v \, \delta(r-R)$, where $v$ is the strength and $R$
is the range, forms the basis of this work in which the transport
properties of a dilute gas of non-relativistic particles are
calculated for densities $n$ for which the average inter-particle
distance $d \sim n^{-1/3} \gg R$.  The delta-shell potential is
particularly intriguing as the scattering length $a_{sl}$
and the effective range $r_0$ can be tuned at will,
including the interesting case of $a_{sl} \rightarrow \infty$ that is
accessible in current atomic physics experiments (see, for e.g., 
Ref. \cite{Kinast05}).
The delta-shell potential also allows us to understand the transport
characteristics of nuclear systems in which the neutron-proton and
neutron-neutron scattering lengths ($a_{sl} \simeq -23.8$ fm and
$-18.5$ fm, respectively~\cite{slengths}) are much longer than the few
fm ranges of strong interactions.  In addition, the
delta-shell potential admits resonances.

For small departures of the distribution function from its
equilibrium, the net flows of mass, energy and momentum are
characterized by the coefficients of diffusion, thermal conductivity
and shear viscosity, respectively. Here we calculate these
coefficients in first and second order of deviations 
(from the equilibrium distribution function) 
using the Chapman-Enskog
method~\cite{Chapman}.  In particular, we examine these transport
properties for varying physical situations including the case of an
infinite scattering length, i.e., the unitary limit, and when
resonances are present.

\section{Two-body interaction}

\begin{figure}[!t]
\includegraphics[width=4.5in]{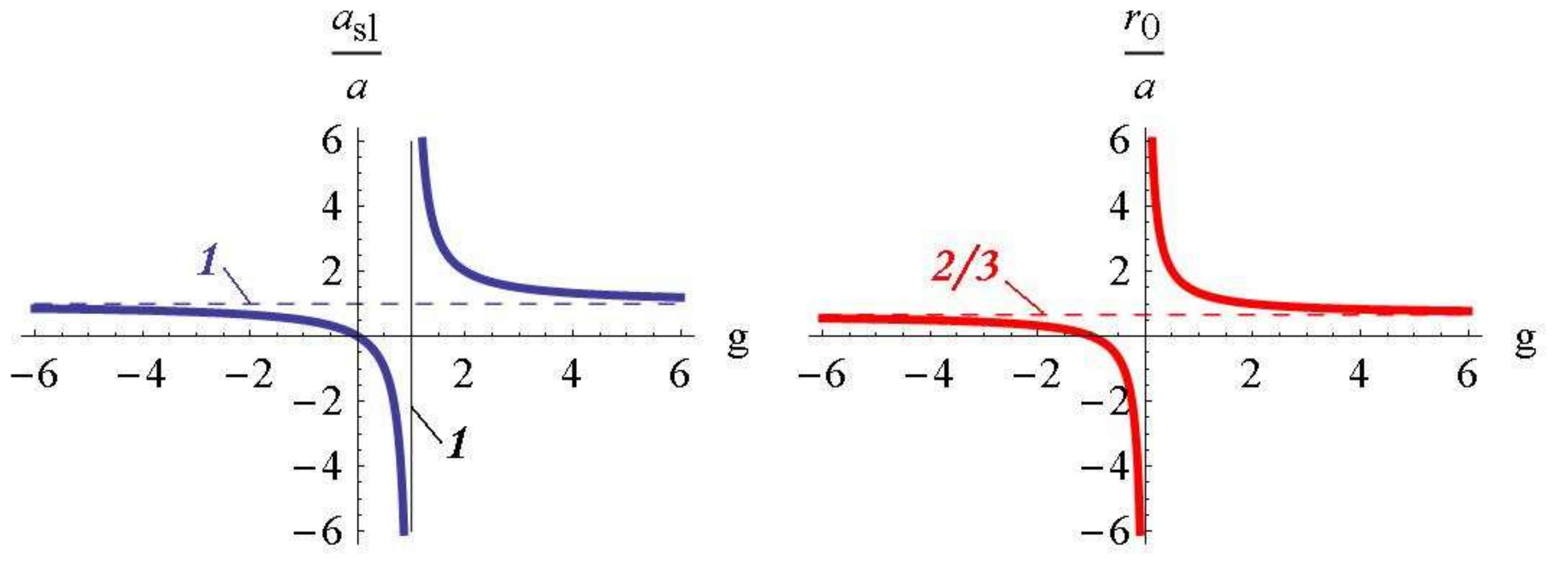}
\caption{Scattering lengths and effective ranges as
  functions of the strength parameter $g$ in Eq.~(\ref{gdef}).}
\label{asl_r0}
\end{figure}

We begin by collecting the two-particle quantum mechanical
inputs~\cite{Gottfried66,Gottfried04,Antoine} and cast them in forms
suitable for the calculation of the transport coefficients.
The Hamiltonian for two particles interacting via the delta-shell potential is
\begin{equation}
\hat H = -\hbar^2 \frac{\Delta}{2\mu}-v \, \delta(r-R)\,,
\end{equation}
where $\mu$ denotes the reduced mass,
$\Delta=\frac{1}{r^2}\frac{\partial}{\partial
r}r^2\frac{\partial}{\partial r}-\frac{\hat L^2}{r^2}$ is the
Laplacian in spherical coordinates ($\hat L$ is the orbital angular
momentum operator), $r$ is the separation distance, $v$ and $R$ are
the strength and range parameters of the potential, respectively. In
terms of the dimensionless variable $\rho=k r$ ($k$ is the wave
number), $E=\frac{\hbar^2 k^2}{2\mu}$, and $l$ the orbital quantum
number, the general solution of Shr\"odinger equation is given by the
spherical Bessel functions $j_l(\rho)$ and $n_l(\rho)$:
\begin{equation}
  \psi(\rho) \equiv {u(\rho)}/{\rho}=A_l j_l(\rho)+B_l n_l(\rho)\,.
\label{general_sln_psi}
\end{equation}
The phase shifts $\delta_l$ are obtained
through~\cite{Gottfried66}
 \begin{equation}
  \tan(\delta_l)=\frac{g \, x \, j_l^2(x)}{1+g \, x \, j_l(x) \, n_l(x)}\,,
\label{tan_dl_gx}
\end{equation}
where $x=k R$, and the single dimensionless parameter
\begin{equation}
\label{gdef}
g=2\mu v \, R/\hbar^2
\end{equation}
controls {\em all} physical outcomes.  At low energies ($l=0$), the
scattering length $a_{sl}$, the effective range $r_0$ and the shape
parameter $P$, are defined through~\cite{Newton}~:
\begin{equation}
  k \, cot(\delta_0) = -\frac{1}{a_{sl}}+r_0\frac{k^2}{2}-P \, 
r_0^3 k^4+O(k^6)\,, \nonumber \\
\end{equation}
\begin{equation}
{a_{sl}} = \frac{R g}{g-1}\,,~~
{r_0} = 
\frac{2 R}{3}\left(1+\frac{1}{g}\right) 
\textrm { and }
P=-\frac{3}{40}\frac{g^2(3+g)}{(1+g)^3}\,. \nonumber
\end{equation}
Note that as $g\rightarrow 1$, $a_{sl}\rightarrow \infty$,
but $r_0/R \rightarrow 4/3$ and $P\rightarrow 3/80$.
For $l > 0$, Newton's generalization of scattering lengths and
effective range parameters reads as~\cite{Newton}
\begin{equation}
k^{2l+1} \, cot(\delta_l) = 
-\frac{1}{a^{(l)}_{sl}}+r^{(l)}_0\frac{k^2}{2}+O(k^4)\,.
\label{asll_r0_P_def}
\end{equation}
For the delta-shell potential, the $l \geq 0$ results are
\begin{equation}
  \frac{a^{(l)}_{sl}}{R^{2l+1}} = 
\frac{(2l+1)}{\left((2l+1)!!\right)^2}\left(\frac{g}{g-(2l+1)}\right)\,,
\label{asll}
\end{equation}
\begin{equation}
  \frac{r^{(l)}_0}{R^{1-2l}} = 
\frac{2 \left((2l+1)!!\right)^2}{(2l+3)(2l-1)}\left(\frac{2l-1}{g}-1\right)\,.
\label{rl_0}
\end{equation}
The scattering lengths in Eq. (\ref{asll}) become very large for
values of $g \rightarrow 2l + 1$.  In Fig. \ref{asl_r0}, we show
$a_{sl}$ and $r_0$ as functions of the strength
parameter $g$ for $l=0$.  The diverging tendencies of of $|a_{sl}|$ as
$g\rightarrow 1$ and $|r_0|$ as $g\rightarrow 0$ are clearly
evident. The case of $g\rightarrow -\infty$ corresponds to the
hard-sphere case.

\section{Transport coefficients}

\par We turn now to calculate the coefficients of diffusion, thermal
conductivity and shear viscosity, based on the Chapman-Enskog
approach~\cite{Chapman,thickbook}.
The central ingredient is the transport cross-section of order $n$: 
\begin{equation}
  \phi^{(n)}=2 \pi \int_{-1}^{+1} d\cos\theta(1-\cos^n\theta) 
\frac{d\sigma(k,\theta)}{d\Omega}\bigr|_{c.m.},
\label{tr_crosssec}
\end{equation}
where the scattering angle $\theta$ and the collisional differential
cross-section $\frac{d\sigma(k,\theta)}{d\Omega}\bigr|_{c.m.}$ are
calculated in the center of mass reference frame of the two colliding
particles. For indistinguishable particles, an expansion of the
cross-section in partial waves $\sum'_{l} (2l+1)(e^{i 2 \delta_l}-1)
P_l(\cos\theta)$ and the orthogonality of the Legendre polynomials
$P_l$ renders the integral above to the infinite sums
\begin{eqnarray}
q^{(1)}\equiv\frac{\phi^{(1)}}{4\pi
R^2}=\frac{2}{x^2}{\sum_{l}}'(2l+1)\sin^2(\delta_l)\,, \nonumber \\
 \label{tr_crosssec_sum1}
q^{(2)}\equiv\frac{\phi^{(2)}}{4\pi R^2} = 
\frac{2}{x^2}{\sum_{l}}'\frac{(l+1)(l+2)}{(2l+3)}
\sin^2(\delta_{l+2}-\delta_l)
 \label{tr_crosssec_sum2}
\end{eqnarray}
The prime on the summation sign indicates the use of even $l$
for Bosons and odd $l$ for Fermions.  In
Eqs.
~(\ref{tr_crosssec_sum2}), the low
energy hard-sphere cross section ($g\rightarrow -\infty$) $4 \pi R^2$
has been used to render the transport cross sections dimensionless. If
the particles possess spin $s$, then the properly symmetrized forms
are:
\begin{equation}
\begin{array}{ll}
q^{(n)}_{(s)} =  \frac{s+1}{2s+1}~q^{(n)}_{Bose} + 
 \frac{s}{2s+1}~q^{(n)}_{Fermi},& \text{for integer } s\,,\\
q^{(n)}_{(s)} =  \frac{s+1}{2s+1}~q^{(n)}_{Fermi} + 
 \frac{s}{2s+1}~q^{(n)}_{Bose},& \text{for half-integer } s\,. \nonumber
\end{array}
\end{equation}
Here, we will present results for the case of spin-1/2 particles only.
The transport coefficients are given in terms of the transport integrals
\begin{equation}
  \omega_\alpha^{(n,t)}(T)\equiv 
\int_0^{\infty}d\gamma \, e^{-\alpha\gamma^2}\gamma^{2t+3}q^{(n)}(x)\,,
\label{tr_ints}
\end{equation}
where $\gamma= \frac{\hbar k}{\sqrt{2 \mu k_B T}}= \frac{x}{\sqrt{2
\pi}}  \left(\frac{\lambda(T)}{R}\right)$ with the thermal de-Broglie
wavelength $\lambda=(2\pi\hbar^2/m \, k_BT)^{1/2}$.

In what follows, the coefficients of self diffusion
$\mathscr{D}$, shear viscosity $\eta$, and thermal conductivity $\kappa$
are normalized to the corresponding hard-sphere-like values 
\begin{equation}
\tilde{\mathscr{D}}=\frac{3\sqrt{2}}{32}\frac{\hbar}{m n R^3 }\,,\quad
\tilde{\eta} =\frac{5 \sqrt {2}}{32}\frac{\hbar}{R^3}\,,~~
\text{and} ~~
\tilde{\kappa}=\frac{75}{64 \sqrt{2}} \frac{\hbar k_B}{m R^3}\,.  
\label{chars}
\end{equation}
In the first order of deviations from the equilibrium distribution
function, the transport coefficients are 
\begin{eqnarray}
\frac{\left[\mathscr{D}\right]_1}{\tilde{\mathscr{D}}} = 
\left(\frac{R}{\lambda(T)}\right)\frac{1}{\omega_{1}^{(1,1)}(T)}\,, \nonumber\\
\frac{\left[\eta \right]_1}{\tilde{\eta}} = 
\frac{\left[\kappa \right]_1}{\tilde{\kappa}} = 
\left(\frac{R}{\lambda(T)}\right)\frac{1}{\omega_{1}^{(2,2)}(T)}\,,
\label{Diffvisq}
\end{eqnarray}
Equation (\ref{Diffvisq}) shows clearly that if $\omega_1^{(2,2)}$ is
$T$-independent (as for hard-spheres with a constant cross section),
the shear viscosity exhibits a $T^{1/2}$ dependence which arises
solely from its inverse dependence with $\lambda (T)$. For
energy-dependent cross sections, however, the temperature dependence
of the viscosity is sensitive also to the temperature dependence of
the omega-integral.

The second order results can be cast as 
\begin{equation}
\frac{\left[\mathscr{C}\right]_2}{\left[\mathscr{C}\right]_1} = 
\left(1+\delta_{\mathscr{C}}(T)\right)
\left(1\pm n \lambda^3 \epsilon_{\mathscr{C}}(T)\right)\,,
\label{Diff2}
\end{equation}
where $\mathscr{C}$ is $\mathscr{D}$ or $\eta$ or $\kappa$, and  the $\pm$
refers to Bose ($+$) and Fermi ($-$) statistics. For example, 
\begin{eqnarray}
\delta_{\eta} &\equiv& 
\frac{3(7\omega_{1}^{(2,2)}-2\omega_{1}^{(2,3)})^2}
{2\left(\omega_{1}^{(2,2)}\left(77\omega_{1}^{(2,2)} + 
6\omega_{1}^{(2,4)}\right)-6\left(\omega_{1}^{(2,3)}\right)^2\right)}\,,
\nonumber \\
\epsilon_{\eta} &\equiv& 
2^{-7/2} \left[4-\frac{128}{3^{3/2}}
\frac{\omega_{4/3}^{(2,2)}}{\omega_{1}^{(2,2)}}\right]\,.
\label{epseta}
\end{eqnarray}
\begin{figure}[!t]
\includegraphics[width=4.5in]{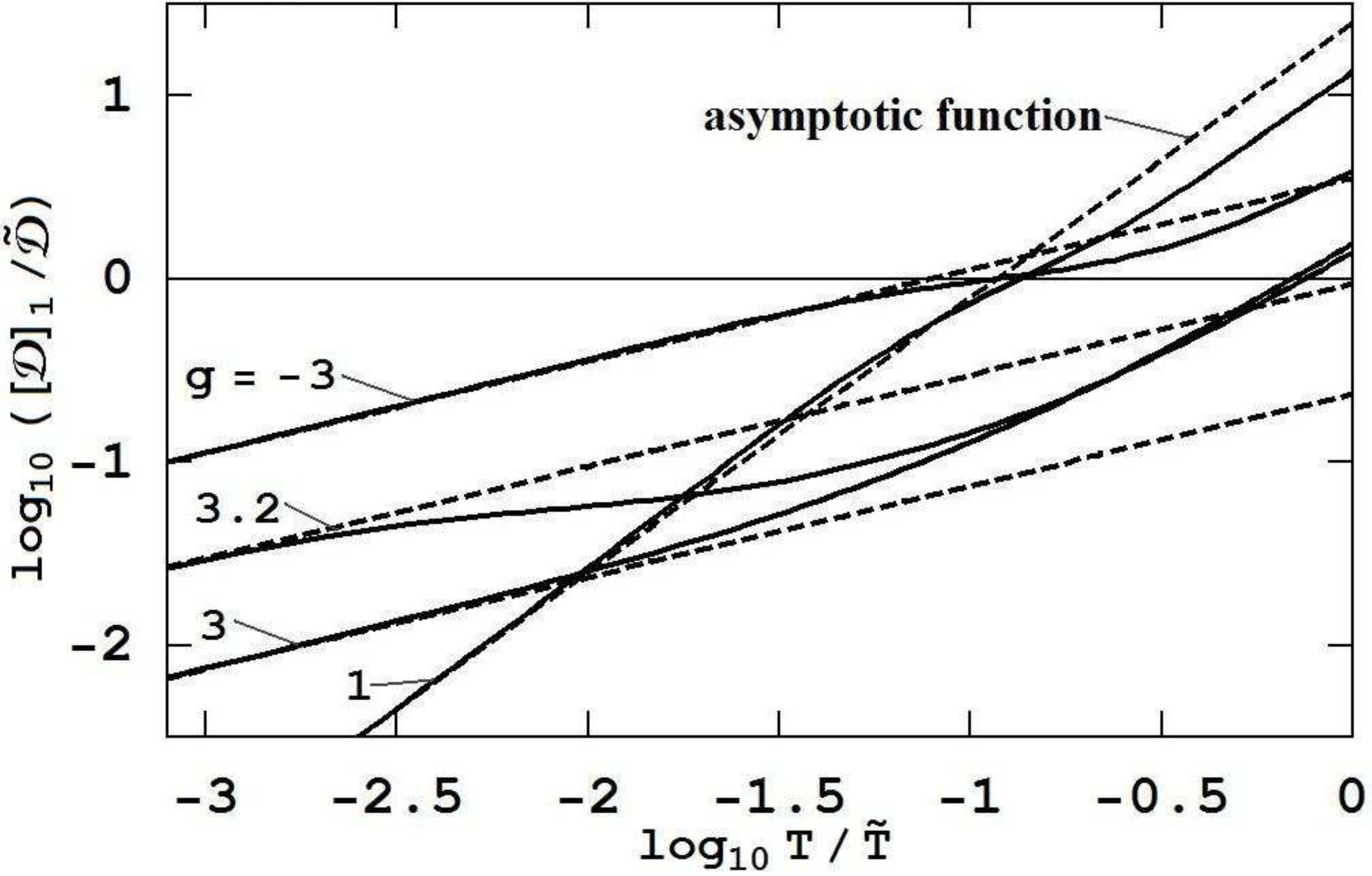}
\caption{Normalized (to $\tilde{\mathscr{D}}$ in Eq.~(\ref{chars}))
diffusion coefficient as a function of normalized (to $\tilde{T}$ in
Eq.~(\ref{Ttilde})) temperature for various values of the strength
parameter $g$ in Eq.~(\ref{gdef}). Solid curves show the diffusion
coefficient in the first approximation, ${\cal D}_1$ in
Eq.~(\ref{Diffvisq}), and the dashed lines are its asymptotic trends for
$T\ll \tilde T$ in Eq.~(\ref{Ttilde}).}
\label{logDlogT}
\end{figure}
\begin{figure}[!t]
\includegraphics[width=4.5in]{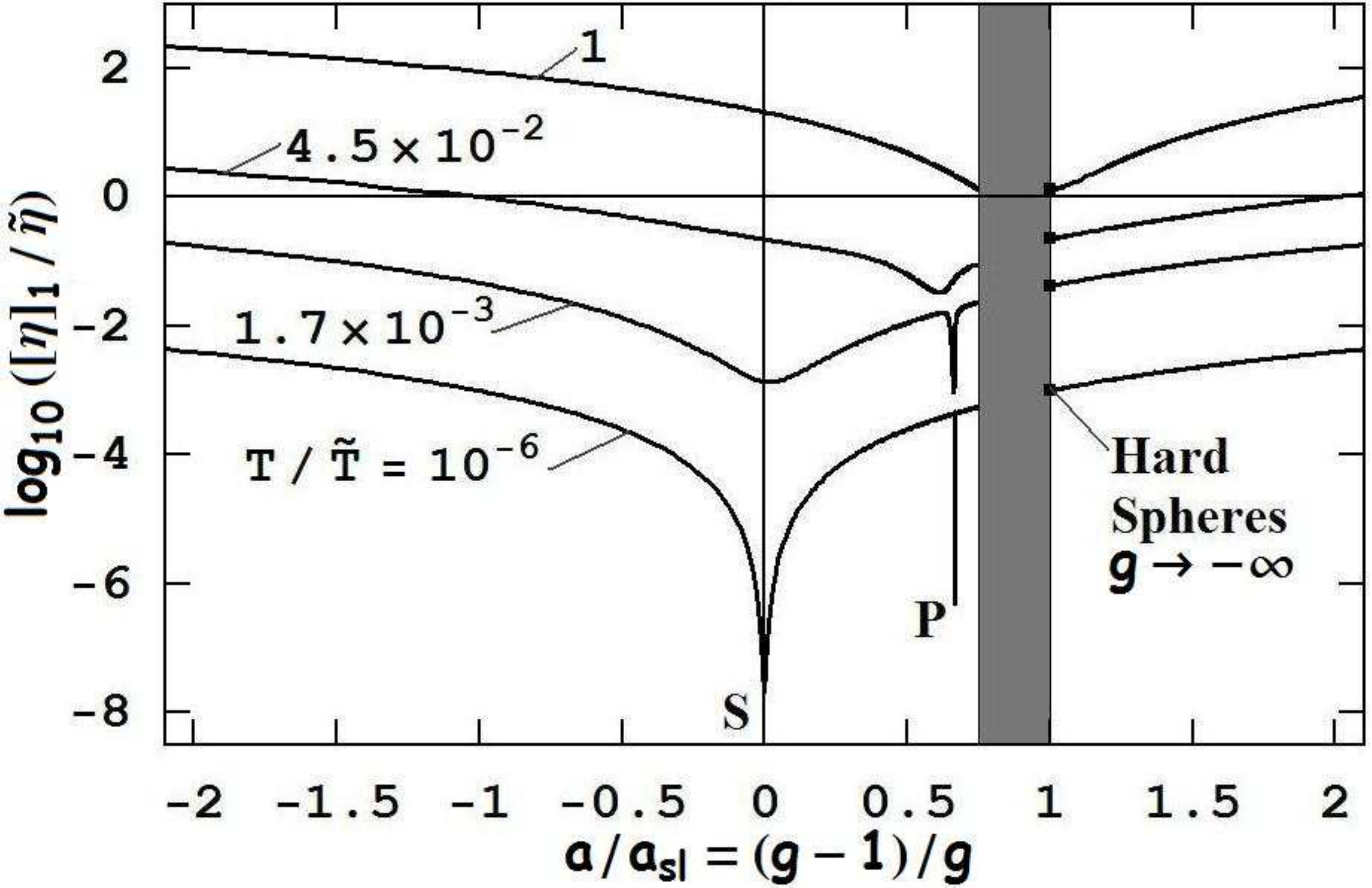}
\caption{The normalized viscosity coefficient in
  Eq.~(\ref{Diffvisq}) as a function of the inverse scattering length.
  Effects due to resonances associated with the partial waves
  $l=0~{\rm and}~1$ are indicated by the letters $S~{\rm and}~P$,
  respectively.  In the vertical shaded region, a large number of
  partial waves are required to obtain convergent results.}
\label{Dislf}
\end{figure}

\section{Asymptotic behavior}

It is useful to define a characteristic temperature
\begin{equation}
\tilde{T}\equiv\frac{2\pi\hbar^2}{k_Bm R^2}~~\text{or}~~
\frac{T}{\tilde{T}}=\left(\frac{R}{\lambda}\right)^2
\label{Ttilde}
\end{equation}
in terms of which limiting forms of the transport coefficients can be
studied.  For spin $1/2$ statistics, Fig. \ref{logDlogT} shows
$\left[\mathscr{D} \right]_1$ from Eq. (\ref{Diffvisq}) normalized to
$\tilde{\mathscr{D}}$ from Eq. (\ref{chars}).  As expected, the
diffusion coefficient grows steadily with temperature. For $g\neq
1,3$, and for $T\ll\tilde{T}$, we find the asymptotic behavior (the
dashed lines in Fig.~\ref{logDlogT})
\begin{equation}
\frac{\mathscr{D}}{\tilde{\mathscr{D}}}\to 2
\left(\frac{1-g}{g}\right)^2\sqrt{\frac{T}{\tilde{T}}}\,.
\label{asDg}
\end{equation}
The cases $g\rightarrow 1,3$ require special consideration.  In these
cases, the asymptotic behavior for $T\ll \tilde T$ is obtained from a
series expansion in $x=k R$ of the transport cross sections
$q^{(n)}(x)$ in Eqs. 
~(\ref{tr_crosssec_sum2}), and subsequent integrations of
Eq.~(\ref{tr_ints}). We find the limiting forms:
\begin{eqnarray}
\frac{\mathscr{D}}{\tilde{\mathscr{D}}} &\to& \left\{ \begin{array}{ll}
8\pi\left(T/{\tilde{T}}\right)^{3/2} &\textrm { for }
g=1 \\
\frac{4}{17}\left(T/{\tilde{T}}\right)^{1/2} &\textrm { for } g=3\,.
\label{asD13}
\end{array}
\right. 
\end{eqnarray}
The asymptotic behavior (for $T \ll \tilde T$) of the 
shear viscosity $\left[\eta \right]_1$ is similar to that of
the diffusion coefficient: 
\begin{eqnarray}
\frac{\eta}{\tilde{\eta}} &\to&
\left\{ \begin{array}{lll}
\left(\frac{1-g}{g}\right)^2 \left(T/{\tilde{T}}\right)^{1/2} &
\textrm { for } g \neq 1,3 \\
6\pi\left(T/{\tilde{T}}\right)^{3/2} &\textrm { for }
g=1 \\
\frac{1}{6}\left(T/{\tilde{T}}\right)^{1/2} &\textrm { for } g=3\,.
\label{asEta13}
\end{array}
\right. 
\end{eqnarray}
It is interesting that even at the two-body level, the coefficients of
diffusion, thermal conductivity and viscosity acquire a significantly
larger temperature dependence as the scattering length 
$a_{sl} \rightarrow \infty~(g \rightarrow 1)$.

\par Figure \ref{Dislf} shows the normalized viscosity coefficient as
a function of $\frac{R}{a_{sl}}=\frac{g-1}{g}$ for $T <
\tilde{T}$. Enhanced cross sections at resonances produce significant
drops in the viscosity as $g\rightarrow 2l+1$. The widths of the dips
in viscosity decrease with increasing values of $l$. For $T\leq
\tilde{T}$, the dips become less prominent and disappear for $T \geq
\tilde T$. For the densities and temperatures considered, the
contribution from the second approximation is small as
$|\delta_{\eta}| < 0.15$.
\begin{figure}[!t]
\includegraphics[width=4.5in]{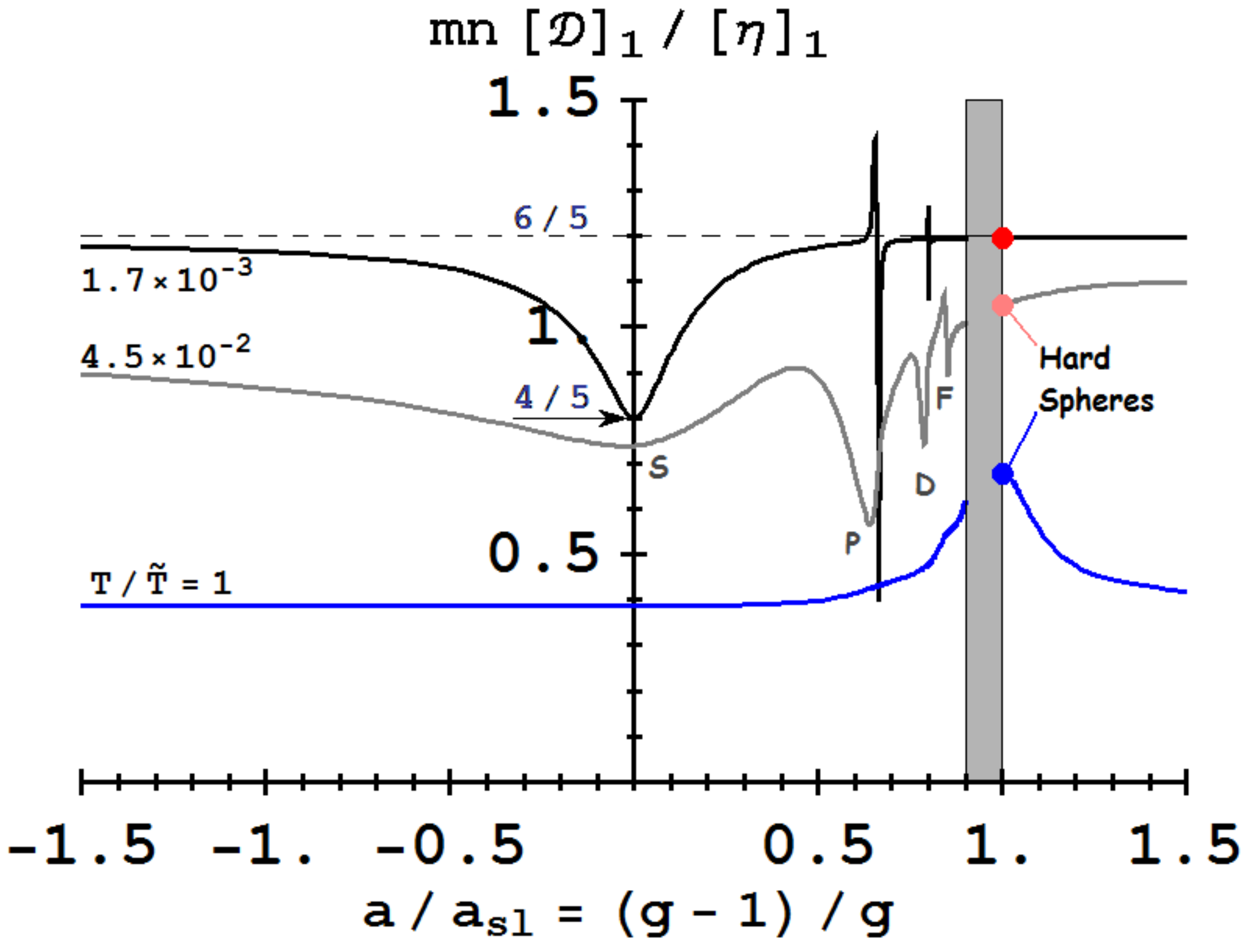}
\caption{Ratio of diffusion (times $mn$) to shear
  viscosity versus inverse scattering length.  In the vertical shaded
  region, a large number of partial waves are required to obtain
  convergent results. The hard-sphere results ($g\to-\infty$) are
  shown by solid circles.}
\label{D1E3}
\end{figure}
\begin{table}[!t]
\caption{First order coefficients of diffusion (times $mn$), shear
viscosity, and their ratios  for $T\ll \tilde{T}$ for select strength
parameters $g$. The unitary limit ($g=1$) result for $\eta$ was
obtained earlier in Ref. \cite{Massignan}.}
\begin{tabular}{@{}cccl@{}} 

\hline \\
\hline \\[-1.8ex] 
$g$ & $mn{\cal D}$  & $\eta$   & $mn{\cal D}/\eta$  \\[0.8ex] 

\hline \\[-1.8ex] 
1 & $  \frac {3{\sqrt 2}\pi}{4} \frac {\hbar}{\lambda^3}$
  & $ \frac {15{\sqrt 2}\pi}{16} \frac {\hbar}{\lambda^3}$ & $ \frac
45 =0.80$\\[0.1in]

\hline \\[-1.8ex] 
3 & $  \frac {3}{68{\sqrt 2}} \frac {\hbar}{\lambda~R^2}$
  & $  \frac {5}{96{\sqrt 2}} \frac {\hbar}{\lambda~R^2}$ &
$ \frac {72}{85}=0.85$\\[0.1in]

\hline \\[-1.8ex] 
$\neq 1,3$  & $  \frac {3{\sqrt 2}}{16} \frac {\hbar}{\lambda~a_{sl}^2}$
  & $  \frac {5{\sqrt 2}}{32} \frac {\hbar}{\lambda~a_{sl}^2}$ &
$ \frac 65 =1.20$\\[1.8ex]

\hline
\hline
\end{tabular}
\label{table1}
\end{table}

\par In Fig.~\ref{D1E3}, the ratio of the coefficients of diffusion
(times $mn$) and viscosity are shown as functions of $R/a_{sl}$ for
$T\leq \tilde T$. As expected, the largest variations in this ratio occur
as $g\rightarrow 2l+1$, that is, as resonances are approached.  For
$T\ll\tilde{T}$, the ratio approaches the asymptotic value $6/5$ away
from resonances and $4/5$ for the $S$-wave resonance (a point). With
increasing $T$, resonances become progressively broader with
diminishing strengths; for $T\approx\tilde{T}$ resonances disappear.
For comparison, this figure also includes results for hard spheres.

The coefficient of viscosity (as also $mn$ times the coefficient of
diffusion) has the dimension of action ($\hbar$) per unit volume.  The
manner in which the effective physical volume $\cal V$ changes as the
strength parameter $g$ is varied is illuminating as our results for
$T\ll \tilde T$ in Table I shows. In the unitary limit ($g=1$), the
relevant volume is ${\cal V}\propto \lambda^3$. For $g=3$, ${\cal V}
\propto \lambda R^2$ (independent of $a_{sl})$, and for $g\neq 1,3$,
${\cal V} \propto \lambda a_{sl}^2$ (independent of $R$).
\begin{figure}[!t]
\includegraphics[width=4.5in]{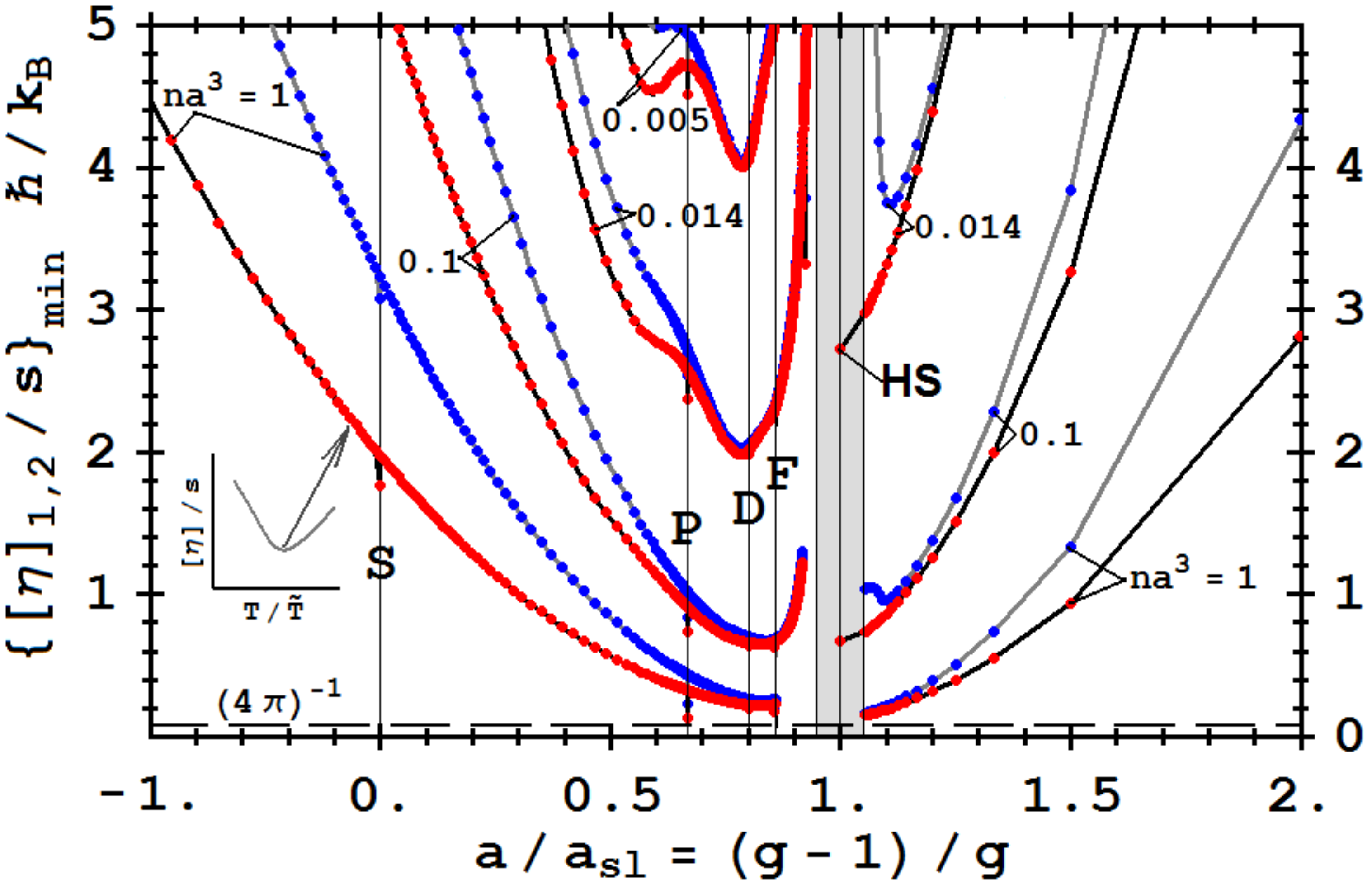}
\caption{The insert shows the occurrence of a
  minimum in the ratio viscosity to entropy density versus $T$.  For
  each $n R^3$, this minimum value in the first (lower curve) and
  second (upper curve) order calculations of shear viscosity is
  plotted versus $R/a_{sl}$ for the indicated $n R^3$.  In the vertical
  shaded region, a large number partial waves are needed.  Vertical
  lines with letters ($S,P,D$, and $F$, respectively) indicate
  resonances associated with the partial waves $l=0,1,2$ and $3$. The
  symbol HS denotes hard-spheres result for which $g \to -
  \infty$. The horizontal dashed line shows the conjectured lower
  bound $1/(4 \pi)$ \cite{Kovtun05}.}
\label{etaOs}
\end{figure}

\section{Ratio of shear viscosity to entropy density}

\par Recently, a lower limit to the ratio of shear viscosity to
entropy density is being sought~\cite{loweretas} with results even 
lower than $(4\pi)^{-1}(\hbar/k_B)$ first proposed in
Ref. X\cite{Kovtun05}. We therefore examine $[\eta]_{1,2}/s$ using the
entropy density
\begin{eqnarray}
s &=& (5/2-\ln(n \lambda^3)+\delta s(T)\, n
R^3)n k_B \,, \nonumber \\
\delta s(T) &=& \left(\frac{a_2(T)}{2} - 
T\frac{da_2(T)}{dT}\right)\left(\frac{\lambda}{R}\right)^3\,,
\label{entropy}
\end{eqnarray}
which includes the second virial correction \cite{thickbook} to the
ideal gas entropy density.  The second virial coefficient \cite{thickbook}
\begin{equation}
\label{a2}
\begin{split}
a_2(T) =& \mp 2^{-5/2}-2^{3/2}{\sum_l}'(2l+1)\\
\times& \left(e^{-E_l/(k_B T)}+\frac{1}{\pi}\int_0^\infty dx\, 
\frac{\partial \delta_l}{\partial x}e^{-\xi(T) x^2}\right)\,,
\end{split}
\end{equation}
where the prime indicates summation over even $l$'s for Bosons ($-$)
and odd $l$'s for Fermions ($+$), $E_l$ is the energy of the bound
state with angular momentum $l$ and $\xi(T) = (\lambda/R)^2/(2 \pi)$.
The insert in Fig.~\ref{etaOs} shows a characteristic minimum of
$\eta/s$ versus $T$ for fixed dilution $n R^3$ and strength $g$.
Values of $[\eta]_{1,2}/s$ at the minimum are also shown functions
of $R/a_{sl}$ for several $n R^3$. The lower (upper)
curve for each $n R^3$ corresponds to the first (second) order
calculations of $\eta$.  The large role of improved estimates of
$\eta$ on the ratio $\eta/s$ is noticeable. The case of $n R^3=1$
possibly requires an adequate treatment of many-body effects not
considered here.  We can, however, conclude that 
in the dilute gas limit $\eta/s$ for the delta-shell gas
remains above $(1/4\pi)\hbar/k_B$.
 
Our analysis here of the transport coefficients of particles subject
to a delta-shell potential has been devoted to the dilute gas
(non-degenerate) limit, in which two-particle interactions dominate,
but with scattering lengths that can take various values including
infinity. Even at the two-body level considered, a rich structure in
the temperature dependence and the effective physical volume
responsible for the overall behavior of the transport coefficients are
evident.  The role of resonances in reducing the transport
coefficients are amply delineated.  Matching our results to those of
intermediate and extreme degeneracies~\cite{Massignan,Rupak07} which
highlight the additional roles of superfluidty and superconductivity
reveals the extent to which many-body effects play a crucial role.

\section*{Personal Note (Madappa Prakash)} 

This contribution to Gerry's Festschrift is in collaboration with my
former graduate student Sergey Postnikov, who obtained his Ph.D degree
in 2009, and who would have impressed Gerry with his originality.  For
Gerry, it is important that researchers be passionate about physics. I
know that Sergey is in the category of people of whom Gerry would have
approved.  Gerry is very fond of schematic models that captures the
essential physics.  Our study here is an offer to the spirit in which
the famous and classic Brown and Bolsterli paper~\cite{BB59} on 
``Dipole State in Nuclei'' was written. I always chide him
with the remark ``You should strive for that excellence''. He always
retorts ``What do you mean?  I always do''.

Gerry cares a lot about both undergraduate and graduate
students. During the years 1981-2005, I have known of many instances
in which he helped students to achieve their career goals, helping
them with guidance in science, as well as with financial support (from
his own purse when funds could not be arranged officially) and with
emotional backing when needed.

I first saw his uncanny abilities with students during a course he
taught with Kevin Bedell in 1982-83 on the applications of Landau
Fermi-liquid theory.  Gerry gave out many ideas, as a result of which,
even before the end of the course, there were three articles written
for Physical Review Letters, which were quickly accepted for
publication.  I was impressed.

To this day, I wish I had Gerry's ways with students. Gerry and I
taught many courses together at Stony Brook.  When we got students'
evaluations, I usually got a numerical score that was better than
Gerry's. However, students invariably said ``Gerry is a fun teacher'',
a compliment that I never got. When we talked about our teaching
skills, he said ``how come I don't get your scores?'' My reply was
``how come I don't get your compliments?''

As Gerry is recovering from his illness, I am now supervising graduate
student Constantinos Constantinou, who began his research with
Gerry. I hope I do as well for Constantinos as Gerry would have done.

\section*{Acknowledgements} 

This research was supported by the Department of Energy under the grant
DE-FG02-93ER40756.

\end{document}